\documentclass{aa}
\usepackage{amsmath} 
\usepackage{graphicx}
\usepackage{natbib}
\bibpunct{(}{)}{;}{a}{}{,}
\usepackage{txfonts}
\newcommand{\cmd}{\,cm$^{-2}$}   
\newcommand{\kms}{\,km\,s$^{-1}$}
\newcommand{\kpc}{{\rm kpc}}
\newcommand{\myr}{\,$M_{\sun}\,{\rm yr}^{-1}$}

\newcommand{\ecsa}{$\rm\,erg\,cm^{-2}\,s^{-1}\,\AA^{-1}$}
\newcommand{\ha}{H$\alpha$}
\newcommand{\hb}{H$\beta$}
\newcommand{\ro}{\,$R_{\sun}$}
\newcommand{\mo}{\,$M_{\sun}$}
\newcommand{\lo}{\,$L_{\sun}$}
\begin{document}
%
%
\title{Measuring the orbital inclination of Z~Andromedae\\
       from Rayleigh scattering}
\author{A.~Skopal \and N.~Shagatova}
\institute{Astronomical Institute, Slovak Academy of Sciences,
        059~60 Tatransk\'{a} Lomnica, Slovakia}
%
\date{Received / Accepted }

\abstract
{
The orbital inclination of the symbiotic prototype Z~And has not 
been established yet. At present, two very different values are 
considered, $i \sim 44^{\circ}$ and $i \ga 73^{\circ}$. 
The correct value of $i$ is a key parameter in, for example, 
modeling the highly-collimated jets of Z~And. 
}
{
To measure the orbital inclination of Z~And. 
}
{
First, we derive the hydrogen column density ($n_{\rm H}$), 
which causes the Rayleigh scattering of the far-UV spectrum 
at the orbital phase $\phi = 0.961\pm 0.018$. 
Second, we calculate $n_{\rm H}$ as a function of $i$ and $\phi$ 
for the ionization structure during the quiescent phase. 
Third, we compare the $n_{\rm H}(i,\phi)$ models with the 
observed value. 
}
{
The most probable shaping of the \ion{H}{i}/\ion{H}{ii} 
boundaries and the uncertainties in the orbital phase limit 
$i$ of Z~And to $59^{\circ} - 2^{\circ}\,/ + 3^{\circ}$. 
Systematic errors given by using different wind 
velocity laws can increase $i$ up to $\sim 74^{\circ}$. 
A high value of $i$ is supported independently by the 
orbitally related variation in the far-UV continuum and the 
obscuration of the \ion{O}{i}]\,$\lambda$1641\,\AA\ emission 
line around the inferior conjunction of the giant. 
}
{
The derived value of the inclination of the Z~And orbital plane 
allows treating satellite components of \ha\ and \hb\ emission 
lines as highly-collimated jets. 
}
\maketitle
\section{Introduction}

Symbiotic stars are long-period interacting binaries comprising 
a cool giant as the donor star and a compact star, most often 
a white dwarf (WD), as the accretor. This composition requires 
large orbital periods, which are typically of a few years. 

The circumstellar environment of symbiotic stars, which consists 
of energetic photons from the hot star and neutral particles from 
the cool giant, represents an ideal medium for the effects of 
Rayleigh and Raman scattering on neutral hydrogen atoms 
\citep[e.g.,][]{nsv89}. 
The Rayleigh scattering produces a strong attenuation of the 
continuum around the Ly-$\alpha$ line in systems with a high 
orbital inclination at positions with the giant in front 
during quiescent phases \citep[][]{inv89}. 
The strength of this attenuation is thus determined by 
$n_{\rm H}$ between the hot star and the observer. 
Its quantity is a function of the binary position and the 
orbital inclination, depending on the ionization structure 
of the binary. For the steady state situation and the quiescent 
phase, the \ion{H}{i}/\ion{H}{ii} boundaries were first 
described by \cite{stb} (hereafter STB). In the model, the 
neutral zone is located symmetrically around the binary axis 
and is usually cone-shaped with the giant at its top facing 
the ionizing source \citep[e.g., Fig.~7 of][]{f-c88}. 

Z~And is the prototype of the symbiotic stars class. The binary 
consists of a late-type, M4.5\,III giant and a WD accreting 
from the giant's wind on the 759-day orbit 
\citep[e.g.,][]{nv89,fekel+00b}. 
Z~And was considered to be a non-eclipsing binary. Based on 
optical observations, \citet{mk96} suggested the orbital 
inclination $i \approx 50^{\circ} - 70^{\circ}$. Based on 
polarimetric measurements of the Raman scattered $\lambda 6830$ 
line, \citet{ss97a} and \cite{isogai+10} determined that 
$i = 47^{\circ}\pm 12^{\circ}$ and 
41$^{\circ}\pm 8^{\circ}$, respectively. 
On the other hand, a high orbital inclination was supported by 
(i) 
the eclipse-like effect measured in the light curve at the 
position of the inferior conjunction of the giant, which 
suggested $i > 76^{\circ}$ \citep[][]{sk03}, 
(ii) 
the polarimetric measurements in the continuum, whose modulation 
along the orbit corresponds to $i = 73^{\circ} \pm 14^{\circ}$ 
\citep[][]{isogai+10}, 
(iii) 
the two-temperature type of the ultraviolet (UV) spectrum that 
develops during active phases of systems with a high orbital 
inclination \citep[][]{sk05}, and 
(iv) 
the presence of the Rayleigh attenuated far-UV continuum measured 
around the inferior conjunction of the giant during the quiescent 
phase \citep[e.g.][]{sk03}. 

Knowledge of the correct value of $i$ is a key parameter in, 
for example, modeling the enhanced bipolar mass outflow, 
launched recently by Z~And \citep[][]{sp06}. 
If $i < 55^{\circ}$, this outflow can be treated as 
a radiatively accelerated wind \citep[][]{tomov+10,kilpio+11}. 
Otherwise, this type of mass outflow represents highly 
collimated jets \citep[e.g.,][]{sk+09}. 

In this paper, we model the effect of the Rayleigh scattering 
measured in the spectrum of Z~And from the quiescent phase to 
derive $i$. 
In Sect.~2, we introduce observations and derive the $n_{\rm H}$ 
parameter, which causes the attenuation effect. 
In Sect.~3, we introduce the ionization structure of the binary 
and calculate $n_{\rm H}$ as a function of the orbital phase 
and the inclination. In Sect.~4, we compare the $n_{\rm H}(i,\phi)$ 
models with the measured value and in this way derive the range 
of $i$ from Rayleigh scattering. The conclusions are found in 
Sect.~5. 
%
%
\begin{figure}
\centering
\begin{center}
\resizebox{\hsize}{!}{\includegraphics[angle=-90]{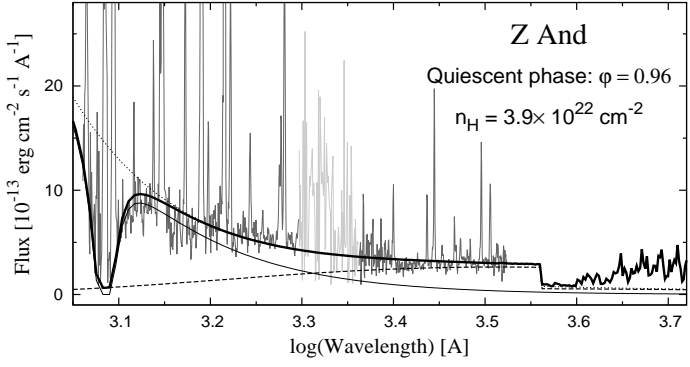}}
\end{center}
\caption[]{UV observations of Z~And carried out close to the 
inferior conjunction of the giant during its quiescent phase 
(on February 3rd 1988). Compared is the model SED (heavy solid line) 
with its components from the nebula (dashed line) and the hot star 
(thin solid line). The strong attenuation of the continuum around 
the Ly-$\alpha$ line is caused by the Rayleigh scattering process. 
Dotted line represents the non-attenuated light (see text for 
details). 
}
\label{fig:sed}
\end{figure}

\section{Observations}

There are two pairs of UV spectra of Z~And in the {\em International 
Ultraviolet Explorer} (\textsl{IUE}) archive taken close to the 
position of the inferior conjunction of the giant 
during the quiescent phase: SWP07410 + LWR06393, from December 
15th 1979, and SWP32845 + LWP12624, from February 3rd 1988. 
Both spectra display signatures of the Rayleigh scattering 
attenuation in the continuum around the Ly-$\alpha$ line. 
However, the former has underexposed the far-UV continuum and 
is affected by a strong geocoronal Ly-$\alpha$ emission. 
Therefore, we treated only the latter spectrum. According to the 
orbital elements of \cite{fekel+00b}, the spectrum was obtained 
at the orbital phase $\phi = 0.961\pm 0.018$. Its spectral 
energy distribution (SED) in the continuum was modeled in the 
same way as by \cite{sk05}. 
Here, we simplified the model by using only one nebula and 
adopting zero He$^{++}$ abundance (Fig.~\ref{fig:sed}). 
The strong depression of the continuum around Ly-$\alpha$ is 
caused by the Rayleigh scattering of the continuum photons on 
the neutral atoms of hydrogen \citep[e.g.,][]{inv89,vo91,d+99}. 
In our case, the depression corresponds to 
$n_{\rm H} = (3.9\pm 0.5)\times 10^{22}$\cmd. 

Having the quantity of $n_{\rm H}$ from observations at 
the given orbital phase, we can calculate its dependence on 
the orbital inclination, $i$, for the ionization structure 
during the quiescent phase. We introduce the model in the 
following section. 

\section{The model}

Here, we determine the ionization structure of a symbiotic binary 
and calculate the column density of neutral hydrogen on the line 
of sight as a function of $\phi$ and $i$. 
Comparing theoretical values of $n_{\rm H}(i,\phi$) with those 
derived independently from observations allows us to estimate 
the inclination of the orbital plane. 

\subsection{$n_{\rm H}$ as a function of the orbital inclination}

\begin{figure}
\centering
\begin{center}
\resizebox{\hsize}{!}{\includegraphics[angle=0]{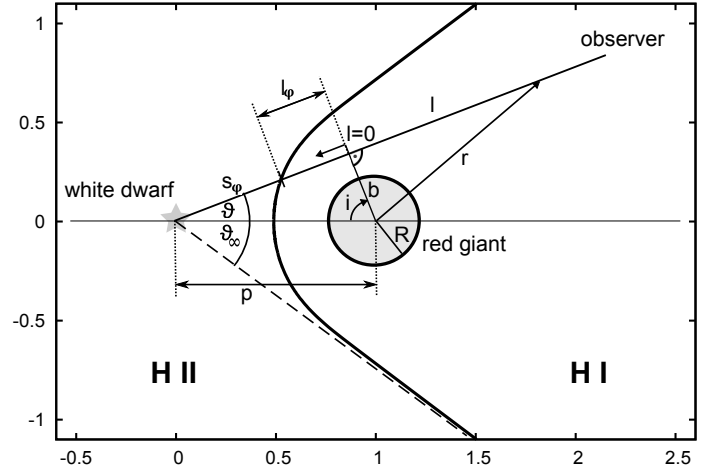}}
\end{center}
\caption[]{
A scheme of a symbiotic binary to calculate $n_{\rm H}$ along 
the line of sight, $l$, throughout the neutral \ion{H}{i} 
region (Eq.~\eqref{nH2}). The thick solid line represents 
the \ion{H}{i}/\ion{H}{ii} boundary (Sect.~3.3) and 
$\vartheta$ is the angle between $l$ and the binary axis. 
Other auxiliary parameters are introduced in Sect.~3.1. 
Distances are in units of the binary components separation, $p$. 
          }
\label{fig:scheme}
\end{figure}

Assuming that the wind from the red giant is spherically symmetric 
and neutral, the continuity equation determines the particle 
concentration of hydrogen as 
\begin{equation}
   N_{\rm H}(r) = 
      \displaystyle\frac{\dot M}{4\pi r^2\mu m_{\rm H}v(r)}, 
\label{NH}
\end{equation}
where $\dot M$ is the mass-loss rate from the giant, $r$ is 
the distance from its center, $\mu$ the mean molecular weight, 
$m_{\rm H}$ the mass of the hydrogen atom, and $v(r)$ velocity 
of the wind. 
Accordingly, we obtain the hydrogen column density, $n_{\rm H}$, 
by integrating Eq.~\eqref{NH} along the line of sight, $l$, 
from the observer ($-\infty$) to the ionization boundary as 
\begin{equation}
 n_{\rm H} = \displaystyle\frac{\dot M}{4\pi \mu m_H}
             \displaystyle\int\limits_{-\infty}^{l_{\varphi}}
             \displaystyle\frac{{\rm d}l}{r^2v(r)},
\label{nH}    
\end{equation}
where $l_\varphi$ is a segment of the line of sight calculated 
from the intersection of its normal, which connects the giant's 
center, to the boundary (Fig.~\ref{fig:scheme}). To determine 
$n_{\rm H}$ we need to know the radius-vector $r$ at each point 
of the line of sight, which is a function of $i$ and the phase 
angle, $\varphi$ ($\varphi=0$ at the giant's inferior conjunction). 
Furthermore, we introduce a parameter $b$ as 
\begin{equation}
   b^2=p^2(\cos^2{i}+\sin^2\varphi\sin^2i),
\label{b2}
\end{equation}
which represents projection of the separation $p$ between
the binary components into the plane perpendicular to the line
of sight (Fig.~\ref{fig:scheme}). Equation~\eqref{nH} may now 
be expressed as 
\begin{equation}
  n_{\rm H} = \displaystyle\frac{\dot M}{4\pi \mu m_H}
              \displaystyle\int\limits_{-\infty}^{l_{\varphi}}
              \displaystyle\frac{{\rm d}l}
                                {(l^2+b^2)v(\sqrt{l^2+b^2})}.
\label{nH2}
\end{equation}
Then, knowing the value of $l_\varphi$ for a given direction to 
the ionizing source and the velocity $v(r)$, we can solve equation 
\eqref{nH2} for $n_{\rm H}$. Figure~\ref{fig:scheme} shows that 
$l_\varphi=\sqrt{p^2-b^2}-s_\varphi$, where $s_\varphi$ is the 
distance from the hot star to the \ion{H}{I}/\ion{H}{II} boundary 
on the line of sight. Its value is given by the ionization 
structure during the quiescent phase (Sect.~3.3), which requires 
knowledge of the velocity $v(r)$ of the giant's wind, which we 
introduce below. 
%
%

\subsection{Velocity profile of the wind from the giant}

\cite{vo91} recognized that the column densities $n_{\rm H}$ 
that he measured from the spectra of EG~And cannot be explained 
by a standard $\beta -$law wind (see, e.g., his Eq.~(6)). The 
observed rapid increase of $n_{\rm H}$ values prior to 
the total eclipse requires a lower velocity in the vicinity 
of the cool giant ($r \approx R_{\rm g}$) and a much 
steeper velocity profile for $0.09 \ga \phi \la 0.14$ 
($r \approx 3 R_{\rm g}$ with respect to that given by 
the $\beta -$law wind \citep[see Figs.~4 and 7 of][]{vo91}. 
Therefore, \citet{vo91} employed Eq.~(4) to reconstruct 
a more appropriate velocity field of the wind from cool 
giants. 

Having many more observations covering the eclipse profile 
of SY~Mus due to the Rayleigh scattering, \citet{pe99} and 
\cite{d+99} solved Eq.~(4) for the velocity law $v(r)$ by 
expanding the function $n_{\rm H}(b)$ into a Taylor series. 
To match the $n_{\rm H}$ values measured around the eclipse, 
both groups of authors confirmed independently the claim 
of a significantly faster acceleration of the wind particles 
from the giant in comparison with the $\beta -$law. 

Re-analyzing the wind acceleration zone for the giant in 
EG~And by using a series of \textsl{FUSE} and \textsl{HST/STIS} 
observations, \cite{crowley+05} recently confirmed that the wind 
acceleration region extends only to $\sim 2.5\,R_{\rm g}$ 
from the limb of the giant. The corresponding wind velocity 
profile was very similar to that derived by \cite{d+99} 
for SY~Mus (see their Fig.~7). 
Therefore, for the purpose of this paper, we use the wind 
velocity law as derived by \cite{d+99}. 
We note that cool giants in both SY~Mus and Z~And have the same 
spectral type \citep[M\,4.5,][]{m+s99} and similar other 
fundamental parameters \citep[see Table~2 of][]{sk05}, which 
also supports adoption of the wind law (\ref{vr}) for the 
case of Z~And. 

According to \cite{d+99}, it was sufficient to approximate the 
$n_{\rm H}(b)$ function by two terms of the Taylor expansion 
($k = 1$ and $k\approx 20)$, which allows expressing the velocity 
law in a form 
%
%
%
\begin{figure}
\centering
\begin{center}
\resizebox{8cm}{!}{\includegraphics[angle=0]{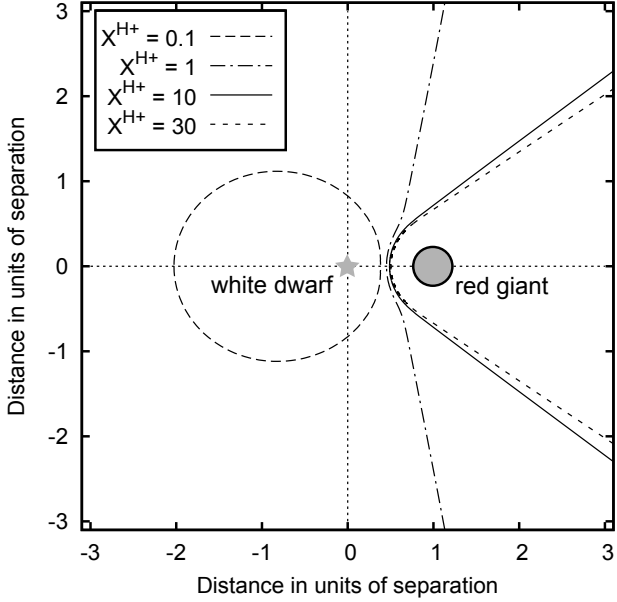}}
\end{center}
\caption[]{
\ion{H}{i}/\ion{H}{ii} boundaries in a symbiotic binary 
given by solution of Eq.~\eqref{XH+=f} for the velocity 
law \eqref{vr} of the giant wind. 
This represents basic ionization structure of the circumstellar 
matter in symbiotic binaries during quiescent phases as 
originally outlined by STB (Sect.~1). 
          }
\label{fig:h1h2}
\end{figure}
%
%
\begin{equation}
  \displaystyle\frac{a}{rv(r)} = 
  \displaystyle\frac{a_1}{\lambda_1r} + 
  \displaystyle\frac{a_k}{\lambda_kr^k}, 
\label{a/rv}
\end{equation}
%
where $a=2\dot M/4\pi\mu m_Hv_\infty R_{\rm g}$, $a_1$, 
$a_k$ and $k$ are parameters, and $\lambda_k$ is defined 
recursively as 
%
%
\begin{equation}
  \lambda_1 = \pi/2, \hskip 1cm 
  \lambda_k=\pi/[(2k-2)\lambda_{k-1}].
\label{lambda}
\end{equation}
%
Fitting the observed $n_{\rm H}$ as a function of the impact 
parameter $b$, \cite{d+99} derived 
 $a_1\approx 4.5\times 10^{23}$\cmd, 
 $a\approx 3\times 10^{23}$\cmd, 
 $k\approx 20$ and 
 $a_{20}\approx 1\times 10^{31}$\cmd. 
Using these values we can rewrite Eq.~\eqref{a/rv} in the form 
%
%
\begin{equation}
   v(r)=\displaystyle\frac{v_\infty}
        {1+\displaystyle\frac{10^8}{3\lambda_{20}r^{19}}}, 
\label{vr}
\end{equation}
%
where $r$ is the distance from the center of the giant in units 
of its radius. According to this law, the main acceleration zone 
of the wind is located between the giant's photosphere and 
$r = 3$ ($v(3)/v_\infty \sim 0.9$, $\lambda_{20} = 0.2838$). 

\subsection{Ionization structure}

The distance $s_\varphi$ from the ionization boundary to the WD 
(Fig.~\ref{fig:scheme}) is determined by a balance between the 
ionizing photons emitted in a small angle around the direction 
$\vartheta$ and recombinations in this angle. According to 
\cite{nv87}, the equilibrium condition can be expressed as 
\begin{equation}
  L_{\rm ph}\displaystyle\frac{\Delta\vartheta}{4\pi}=
      \displaystyle\frac{\Delta\vartheta}{4\pi}\displaystyle
      \int\limits_{0}^{s_\varphi}N_{H^+}(s)N_e(s)
      \alpha_B({\rm H},T_e)4\pi s^2ds,
\label{i=r}   
\end{equation}
where $L_{\rm ph}$ is the number of photons capable of ionizing 
hydrogen that are emitted spherically-symmetrically from the hot 
star per second, $s$ is the distance from the hot star, $N_{H^+}$ 
and $N_e$ are concentrations of protons and electrons, respectively, 
and $\alpha_{\rm B}$ is the total hydrogen recombination coefficient 
for recombinations other than to the ground state. In the $H^+$ 
region, we consider that $N_e = N_H = N_{H^+}$. Then, assuming that 
$T_e$ is constant throughout the nebula and thus also 
$\alpha_{\rm B}$, condition \eqref{i=r} can be expressed as 
\begin{equation}
  f(r,\vartheta) = X^{H+},
\label{XH+=f} 
\end{equation}
where $X^{H+}$ is the ionization parameter (STB), 
\begin{equation}
   X^{H+} = \frac{4\pi (\mu m_H)^2}
                 {\alpha_{\rm B}({\rm H},T_{\rm e})} p L_{\rm ph} 
          \left(\frac{v_\infty}{\dot M}\right)^2, 
\label{XH+}
\end{equation}
and 
\begin{equation}
f(r,\vartheta)=
   p v_\infty^2\displaystyle\int\limits_{0}^{s_\varphi}
    \displaystyle\frac{s^2}{r^4\, v^2(r)}\, {\rm d}s, 
\label{fr}
\end{equation}
where we used the particle concentration in the giant's wind 
as given by Eq.~\eqref{NH}, and
\begin{equation}
   r^2 = s^2+p^2-2sp\cos\vartheta.
\label{r2}
\end{equation}
Examples of \ion{H}{i}/\ion{H}{ii} boundaries are depicted 
in Fig.~\ref{fig:h1h2}. 

\section{Orbital inclination of Z~And}

Having defined ionization structure, we can solve 
Eq.~\eqref{XH+=f} for $s_\varphi$, which determines 
$l_\varphi$ and thus $n_{\rm H}$ 
(see Fig.~\ref{fig:scheme}, Eq.~(4)) for the ionization 
parameter $X^{H+}$. Its value is given by the binary parameters. 
Therefore, first we find a possible range of $X^{H+}$. 

For $P_{\rm orb}$ = 759 days \citep[][]{fekel+00b} and the total 
mass of the binary, 2.6\mo\ \citep[][]{mk96}, we get the 
separation of the stars, $p = 480$\ro. 
Furthermore, we adopt $\alpha_{\rm B}({\rm H},T_{\rm e}) = 
                  1.43\times 10^{-13}$cm$^{3}$s$^{-1}$ 
for $T_e = 20\,000K$ \citep[][]{nv87} and 
$\dot M \sim 7\times 10^{-7}$\myr, derived from the total 
emission measure during quiescence \citep[][]{sk05}. 
This value of $\dot M$ represents an upper limit because 
of the presence of other sources of the nebular radiation, 
e.g., the hot star wind. The hot star luminosity 
$L_{\rm h} \sim 2\,300(d/1.5\kpc)^2$\lo\ and a temperature 
of $\sim$120\,000\,K \citep[Table~3 of][]{sk05} yield 
$L_{\rm ph} \sim 1.5\times 10^{47}(d/1.5\,\kpc)^2$\,s$^{-1}$. 
Assuming a typical value for the terminal velocity for the giant 
wind, $v_\infty \approx 40$\kms\ \citep[][]{reimers}, 
we obtain $X^{H+} \approx 20$. We note that \cite{f-c88} derived 
$X^{H+} \approx 14$. 
However, there are more sources of uncertainties in determining 
the parameter $X^{H+}$: 
(i) 
Terminal velocity for the massive slow wind from the giant 
was also assumed to be of 20\kms\ only \citep[e.g.,][]{d+99}. 
This value scales the above derived $X^{H+}$ with a factor 
of 0.25. 
(ii) 
A smaller distance to Z~And of 1.12\,kpc and 1.2\,kpc 
\citep[][]{f-c88,sok+06} reduces the luminosity and thus 
the value of $X^{H+}$ with a factor of 0.55 and 0.64, 
respectively. 
(iii) 
A maximum Zanstra temperature of 130\,000\,K \citep[][]{m+91} 
gives $L_{\rm h} = 3\,400(d/1.5\kpc)^2$\lo, 
i.e. $L_{\rm ph} = 2.2\times 10^{47}(d/1.5\kpc)^2$s$^{-1}$, 
which enlarges the above value of $X^{H+} = 20$ with a factor 
of $\sim$1.5. 
According to these possibilities we consider a range of 
the ionization parameter as 
\begin{equation}
  3 < X^{H+} \la 30 .
\label{eq:xrange}
\end{equation}
In the following subsection, we explore the $n_{\rm H}(i,\phi)$ 
function at the position of $\phi = 0.961 \pm 0.018$, 
when the \textsl{IUE} observations were exposed. 
%
%
\begin{figure}
\centering
\begin{center}
\resizebox{\hsize}{!}{\includegraphics[angle=0]{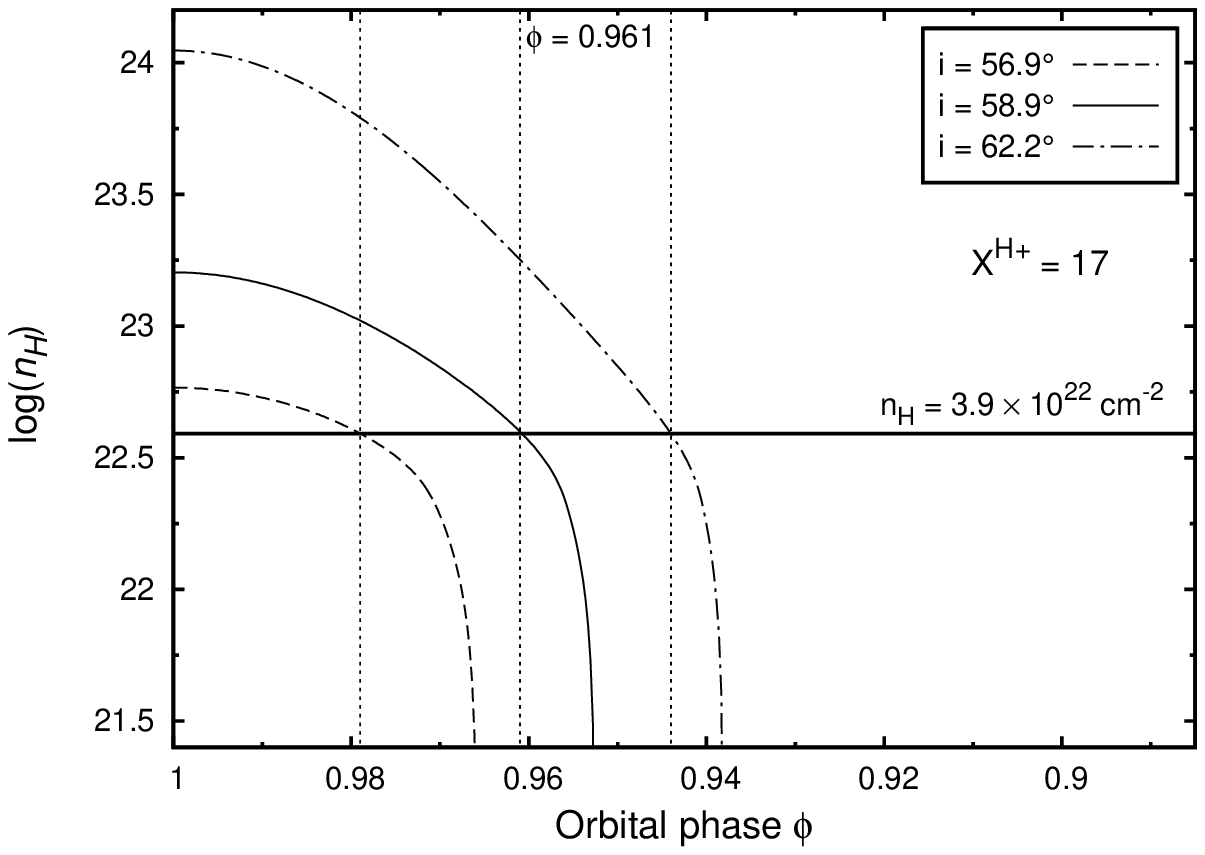}}
\resizebox{\hsize}{!}{\includegraphics[angle=0]{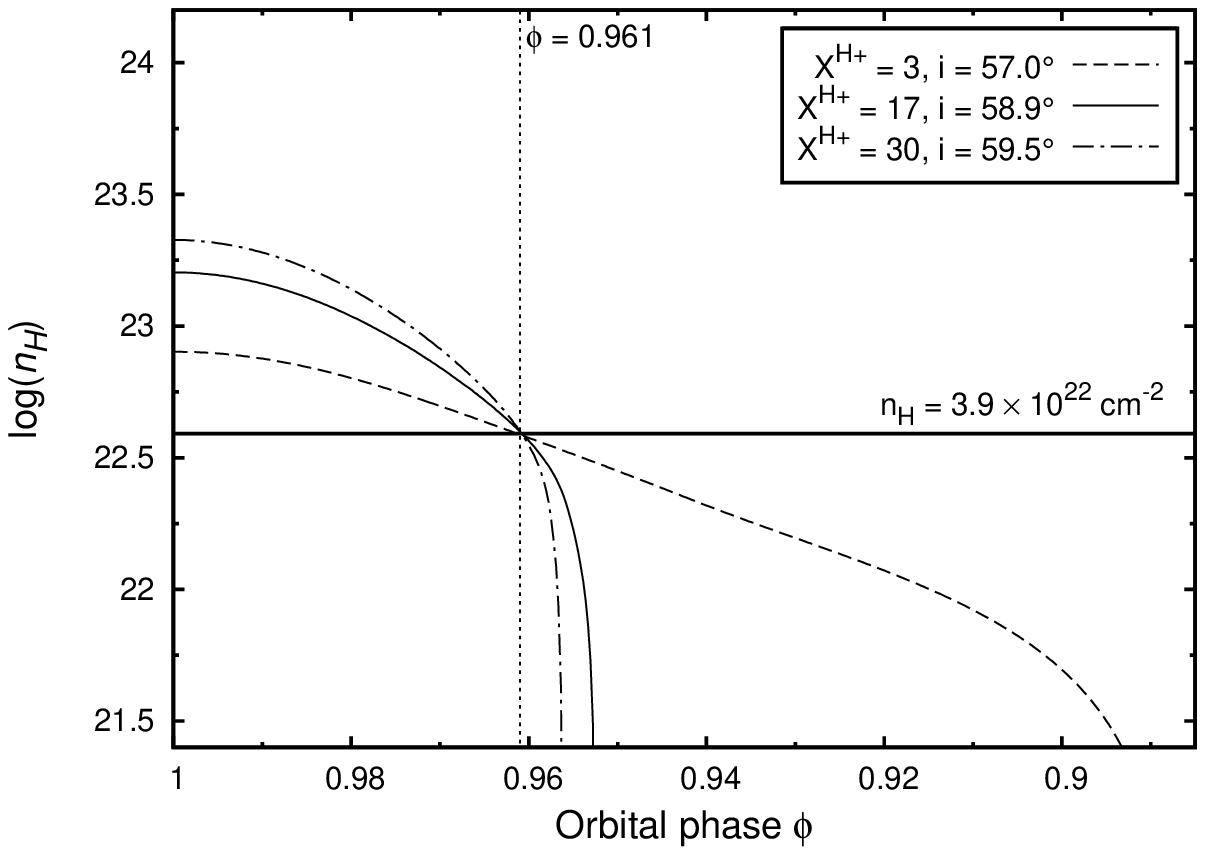}}
\end{center}
\caption[]{
Top: Calculated column densities (Eq.~(4)) as a function of the 
orbital phase for the \ion{H}{i} zone defined by the mean 
value of $X^{H+}=17$ (relations (10) and (13)). 
Particle concentration along the line of sight (Eq.~(1)) was 
calculated for $v_{\infty} = 40$\kms. 
Solutions giving $n_{\rm H} = 3.9\times 10^{22}$\cmd\ 
at the orbital phases $\phi = 0.961 \pm 0.018$ (dotted 
lines) are plotted. 
Bottom: As in the top, but for $3 \lesssim X^{H+} \lesssim 30$ 
and $\phi = 0.961$. 
          }
\label{fig:nhfi}
\end{figure}
%
%
\begin{figure}
\centering
\begin{center}
\resizebox{\hsize}{!}{\includegraphics[angle=0]{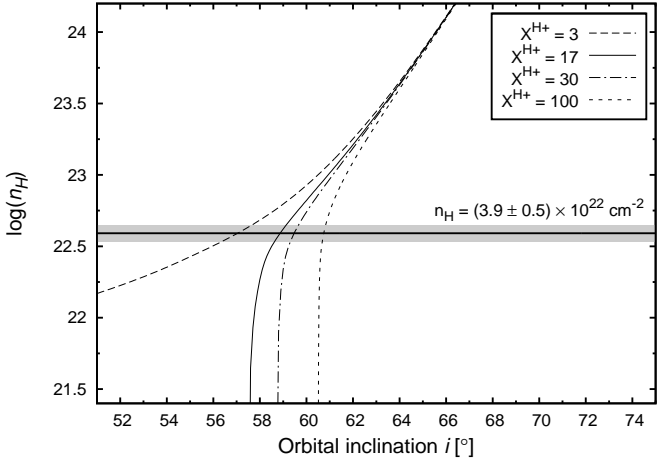}}
\end{center}
\caption[]{
As in the bottom panel of Fig.~\ref{fig:nhfi}, but for 
$n_{\rm H}(i,\phi=0.961)$ as a function of $i$. 
Solution for $X^{H+} = 100$ was added to demonstrate 
the small change of $i$ for very high $X^{H+}$ (Sect.~4.1). 
Intersections of the model curves with the horizontal line 
at the observed value of $n_{\rm H}=3.9\times 10^{22}$\cmd\ 
correspond to $i = 57,~59,~59.5,~{\rm and}~60.^{\circ}7$. 
The shadow  band represents the uncertainty in the measured 
$n_{\rm H}$. 
          }
\label{fig:nhi}
\end{figure}

\subsection{Modeling the column density at $\phi = 0.961$}

The top panel of Fig.~\ref{fig:nhfi} shows the resulting column 
densities as a function of the orbital phase, calculated for 
the average value of $X^{H+} = 17$, for which 
$n_{\rm H}(i,\phi = 0.961 \pm 0.018) = 3.9\times 10^{22}$\cmd. 
The range of possible orbital phases limits the 
inclination angle to 
       $i = 59^{\circ} - 2^{\circ}\,/ + 3^{\circ}$. 
In the bottom panel of Fig.~\ref{fig:nhfi}, the 
$n_{\rm H}(i,\phi)$ function was calculated for the range 
of possible $X^{H+}$ (Eq.~\eqref{eq:xrange}). 
Here, the models for different values of $X^{H+}$, satisfying 
$n_{\rm H}(i,0.961) = 3.9\times 10^{22}$\cmd, restrict 
the orbital inclination to 
       $i = 59^{\circ} - 2^{\circ}\,/ + 1^{\circ}$. 
Small uncertainties reflect small differences between opening 
angles of \ion{H}{i} zones for larger values of $X^{H+}$ 
(see Fig.~\ref{fig:h1h2}). 
Figure~\ref{fig:nhi} illustrates the calculated 
$n_{\rm H}(i,0.961)$ as a function of $i$ for the range of 
the parameter $X^{H+}$. 
From the figure, we can see that the uncertainty in the measured 
value of $n_{\rm H}$ can enlarge the resulting uncertainty 
in $i$ by a few degrees. 
However, systematic errors can result from using a different 
wind velocity law. 
Although the used $v(r)$ satisfies best the measured $n_{\rm H}$ 
as a function of the parameter $b$ (Sect.~3.2), its high values 
$\ga 10^{24}$\cmd\ in the vicinity of the stellar disk 
of the giant cannot be measured accurately 
\citep[see Fig.~5 of][]{d+99}. This precludes a correct 
determination of the wind law. The $n_{\rm H}(b)$ values, as 
measured prior to and after the inferior conjunction of the 
giant, also reflect an asymmetry of the neutral wind zone with 
respect to the binary axis \citep[see Fig.~4 of][]{d+99}.
The used wind law (\ref{vr}) was derived from the egress
data. Thus, the steeper ingress data will correspond to 
a steeper $v(r)$. Therefore, we also investigated a limiting 
case for a maximum steepness of $v(r) = v_{\infty}$. This 
approach yielded 
$i = 74^{\circ}.7 - 4^{\circ}.0\,/ + 10^{\circ}.3$. 
From this point of view, the orbital inclination of Z~And 
can be expected to be in the range of 
$\sim 59^{\circ} - 74^{\circ}$. 

Finally, the dependence of $i$ on $\dot M$ follows that on 
$X^{H+}$, because $X^{H+} \propto \dot M^{-2}$ 
(Eq.~\eqref{XH+}). This implies that 
$\dot M <$ upper limit of $7\times 10^{-7}$\myr\ (see above) 
yields higher $X^{H+}$, which, however, has no significant 
effect on the corresponding value of $i$, because the opening 
angle of the neutral zone, $\vartheta_{\infty}$, decreases 
very slowly for $X^{H+} > 10$ (see Fig.~\ref{fig:h1h2}). 
For example, the lower value of 
$\dot M = 3.1\times 10^{-7}(d/1.5\kpc)^{3/2}$\myr, derived 
by \cite{f-c88} from radio observations, enlarges the range 
of $X^{H+}$ (\ref{eq:xrange}) by a factor of $\sim$5, i.e. 
$15 < X^{H+} \la 150$. Figure~\ref{fig:nhi} also shows the 
model $n_{\rm H}(i,0.961)$ for $X^{H+} = 100$, which 
demonstrates only a small increase of $i$ to 60.$^{\circ}$7. 

\subsection{The far-UV light curve}

\begin{figure}
\centering
\begin{center}
\resizebox{\hsize}{!}{\includegraphics[angle=-90]{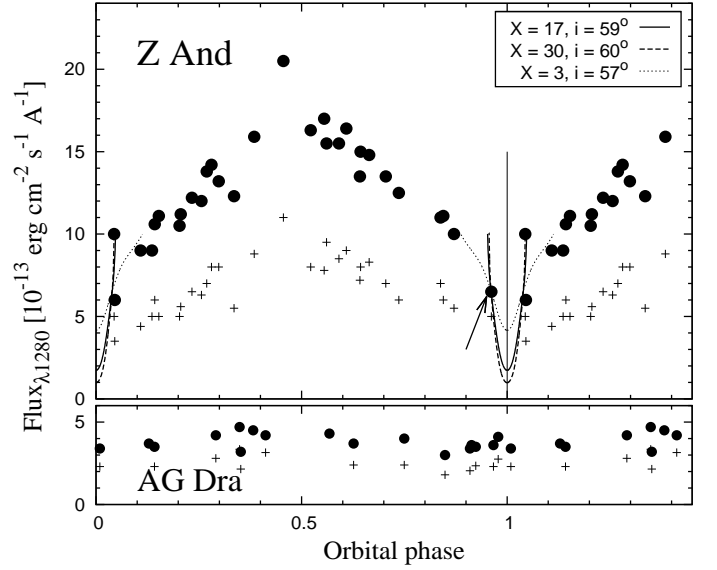}}
\end{center}
\caption[]{
Variation in the far-UV, 1\,280\,\AA\ ($\bullet$) and 
1\,600\,\AA\ ($+$) fluxes of Z~And and AG~Dra along their 
orbits as measured on the \textsl{IUE} spectra during their 
quiescent phases. The arrow points the flux 
from the spectrum in Fig.~\ref{fig:sed}. The lines around 
$\phi = 0~{\rm and}~1$ represent Rayleigh attenuated light 
curves for models in keys. Orbital elements of \cite{fekel+00b} 
were used. 
          }
\label{fig:uvlc}
\end{figure}

The high orbital inclination of Z~And is also supported by 
a large variation in the far-UV continuum along the orbit as 
measured on \textsl{IUE} spectra. Figure~\ref{fig:uvlc} shows 
examples of fluxes measured around $\lambda$1280\,\AA, which 
is relatively free of emission/absorption features. 
Similar variations are indicated also at longer wavelengths, 
where the influence of the Rayleigh scattering is negligible. 
For example, we selected fluxes at $\sim \lambda$1600\,\AA, 
where instead the iron curtain absorption could be pronounced 
\citep[see][]{sa93}. The variation here has a smaller 
amplitude with a shallow minimum with respect to that at 
$\sim \lambda$1280\,\AA. 
Generally, the UV continuum peaks around the superior 
conjunction of the giant, while at the opposite site the 
fluxes are significantly fainter. 
This variability reflects the presence of an additional source 
of extinction, which attenuates more the short wavelength part 
of the UV spectra. This effect is known for eclipsing symbiotic 
binaries RW~Hya, SY~Mus, and EG~And \citep[][]{d+99,crowley}. 
The influence of the absorbing medium was modeled by \cite{ho+94}, 
who found that the absorbing gas produces a modest optical depth 
in the Balmer and Pashen continuum, which can make it 
significantly lower in the UV (see their Fig.~8). 
As the absorbing material is concentrated more on the orbital 
plane and enhanced at positions with the giant in front, 
the large amplitude of the far-UV continuum can be 
attributed to a high orbital inclination of Z~And. 
In contrast, the non-eclipsing symbiotic binary AG~Dra 
\citep[$i\approx 30-45^{\circ}$ or $i\sim 60^{\circ}$,][]
{mika+95,ss97b} does not show such variability 
in the UV spectrum (Fig.~\ref{fig:uvlc}). 

Finally, we calculated the Rayleigh scattered light curves 
for the $n_{\rm H}(i,\phi)$ functions depicted in the 
bottom panel of Fig.~\ref{fig:nhfi}, scaled to the observed 
flux at $\phi = 0.961$ (see Fig.~\ref{fig:uvlc}). 
The aim here is to model the expected Rayleigh attenuation 
as a function of $\phi$ and to check how it is connected 
with the neighboring measurements. 
Assuming that the Rayleigh scattering is the only process 
attenuating the far-UV continuum, we can write the observed 
flux as 
\begin{equation}
  F_{\lambda}(\phi) = F_{\lambda}^{0}\,
             e^{-\sigma_{\lambda}^{\rm Ray} n_{\rm H}(i,\phi)}, 
\label{eq:fray}
\end{equation}
where $F_{\lambda}^{0}$ is the original flux and 
$\sigma_{\lambda}^{\rm Ray}$ is the Rayleigh scattering 
cross section. 
For $F_{1280}(0.961) = 6.5 \times 10^{-13}$\ecsa\ 
(Fig.~\ref{fig:uvlc}), 
$n_{\rm H}(0.961) = 3.9\times 10^{22}$\cmd\ 
(Fig.~\ref{fig:sed})
and 
$\sigma_{1280}^{\rm Ray} = 1.1 \times 10^{-23}$\,cm$^2$ 
\citep[][]{nsv89} 
we get the unscattered flux 
$F_{1280}^{0} = 1.0 \times 10^{-12}$\ecsa. 
Then, according to Eq.~\eqref{eq:fray}, we calculated 
the light curve $F_{1280}(\phi)$ for the three models 
$n_{\rm H}(i,\phi)$ plotted in Fig.~\ref{fig:nhfi}. 
The result is shown in Fig.~\ref{fig:uvlc}. 
The rather narrow ($\la 0.1\,P_{\rm orb}$) and deep model 
light curves reflect a close cut of the \ion{H}{i} region 
for the resulting inclinations and the strong dependence 
of the Rayleigh attenuation on $n_{\rm H}$. 
In other words, due to a small opening angle of the neutral 
zone ($\vartheta_{\infty} \sim 35^{\circ}$ and 34$^{\circ}$ for 
$X = 17$ and $X = 30$, respectively), their cut with the line 
connecting the hot star and the observer along the orbit 
covers only a small segment of the orbital period. 
Overall, the Rayleigh scattering light curves, calculated 
according to our resulting $n_{\rm H}(i,\phi)$ models, seem 
to be consistent with the fluxes measured around the inferior 
conjunction of the giant. 
%
%
%

%
\begin{figure}
\centering
\begin{center}
\resizebox{\hsize}{!}{\includegraphics[angle=-90]{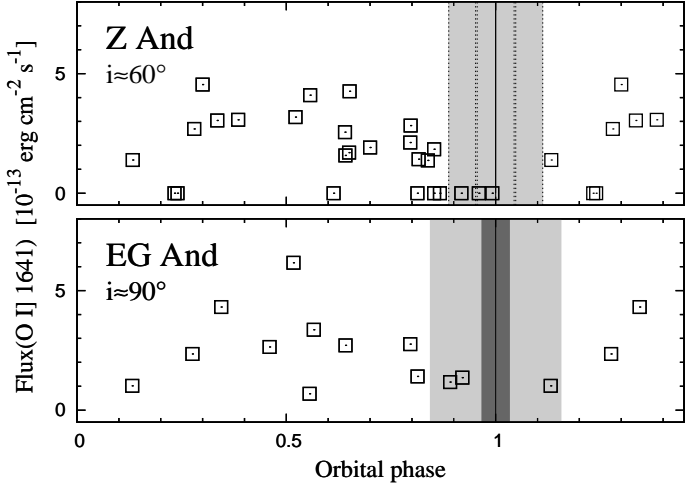}}
\end{center}
\caption[]{
Variations of the \ion{O}{i}]\,1641\,\AA\ line fluxes during 
quiescent phases of Z~And and EG~And as a function of the 
orbital phase according to the ephemeris of \cite{fekel+00a} 
and \cite{fekel+00b}, respectively. The light gray belts 
correspond to the intersection of the \ion{H}{i} zone with 
the line of sight along the orbit (see Sect.~4.3). 
The dark belt reflects a time interval of the total eclipse 
of the point source by the giant in EG~And 
\citep[$\Delta\phi = \pm 0.038$; see Table~1 of][]{crowley}. 
Fluxes were taken from Figs.~4 and 8 of \cite{s+w10}. 
          }
\label{fig:oi]}
\end{figure}

\subsection{A probe of a high inclination with 
            the \ion{O}{i}${\rm ]}$\,1641\,\AA\ line}

\cite{s+w10} investigated observational properties of the 
intercombination transition \ion{O}{i}]\,1641\,\AA\ to 
a metastable state in the spectra of some symbiotic stars 
with the aim of diagnosing the neutral red giant wind. 
Their idea is based on the following: 
(i) the ionization potential of both oxygen and hydrogen is 
comparable, 
(ii) the oxygen resonance multiplet (1302, 1304, 1305\,\AA) is, 
in addition to the ground state, connected to two metastable 
states through emission at 1641\,\AA\ and 2324\,\AA\ 
\citep[see Fig.~1 of][]{s+w10}, and 
(iii) the intercombination transition at $\lambda$1641\,\AA\ 
can be stimulated by absorption from the ground state of the 
\ion{O}{i} 1302\,\AA\ resonance line in the surrounding 
neutral gas. 
This implies that the \ion{O}{i} 1302\,\AA\ emission line, which 
is formed by recombinations within the extended \ion{H}{ii} 
region, surrounds closely the neutral \ion{H}{i} region during 
quiescent phases of symbiotic stars and thus gives rise to 
the measurable \ion{O}{i}]\,1641\,\AA\ emission within the 
neutral wind from the cool giant. 
\cite{s+w10} found that the \ion{O}{i}] variations are strongly 
correlated with the optical light curves and also with the 
\ion{O}{vi} Raman emission at $\lambda$6825, 7082\,\AA. Both 
the \ion{O}{i}] 1641\,\AA\ and the Raman lines are optically 
thin; the transitions are not repeatable: once they happen, 
their photons escape the medium. As a result, the transitions 
arise in the red giant wind close to the \ion{H}{i}/\ion{H}{ii} 
interface, preferentially between the binary components with 
the highest density of the giant's wind (in contrast to the 
Rayleigh scattering, which is a repeatable process and thus 
measures the total thickness of the neutral wind). 
\cite{s+w10} confirmed this view for the eclipsing system 
EG~And, where the \ion{O}{i}] line was obscured around the 
inferior conjunctions of the giant. For low-inclination systems, 
the authors suggested that these lines should always be visible. 

Accordingly, to probe the orbital inclination of Z~And, we 
plotted the \ion{O}{i}]\,1641\,\AA\ line fluxes of Z~And from 
its quiescent phase together with those of the eclipsing 
symbiotic binary EG~And in the phase diagram 
(Fig.~\ref{fig:oi]}). 
The figure demonstrates that around the inferior conjunction 
of the giant ($\phi$ = 0 or 1), the \ion{O}{i}] fluxes are very 
weak (EG~And) or not measurable (Z~And). Maximum values were 
measured when the line of sight does not intersect the neutral 
hydrogen zone (the light gray belts in Fig.~\ref{fig:oi]}). 
For EG~And, the extent of the \ion{H}{i} region in the phase 
diagram ($\Delta\phi = \pm 0.16$) was determined for 
$i = 90^{\circ}$ \citep[][]{v+92} and the position of the 
\ion{H}{i}/\ion{H}{ii} boundary at $b \approx 3.5\,R_{\rm g}$ 
\citep[see Fig.~7 of][ and Eq.~\eqref{b2} here]{crowley+05}. 
In the case of Z~And, the range of intersections of the line 
of sight with the neutral zone in the phase diagram 
($\Delta\phi = \pm 0.11, 0.048, 0.044$; dotted lines in 
Fig.~\ref{fig:oi]}) corresponds to our models for 
$X$ = 3, 17, 30 from the bottom panel of Fig.~\ref{fig:nhfi}. 
The results are consistent with the properties and the formation 
region of the \ion{O}{i}]\,1641\,\AA\ line during quiescent 
phases, as described above. Thus, the occultation of the 
\ion{O}{i}] fluxes of Z~And during its quiescent phase 
independently supports a higher inclination of its orbit. 

\section{Conclusion}

We used \textsl{IUE} observations of the symbiotic star Z~And, 
which were taken during its quiescent phase around the inferior 
conjunction of the giant, to determine the inclination of its 
orbital plane. 

First, we derived $n_{\rm H} = 3.9\pm 0.5 \times 10^{22}$\cmd\ 
(Fig.~\ref{fig:sed}) on the line of sight, which passes throughout 
the neutral zone of the giant's wind towards the hot star at the 
time of the observation (at $\phi = 0.961\pm 0.018$), by modeling 
the Rayleigh scattering of the far-UV continuum (Sect.~2). 
Second, we determined the STB ionization structure for 
the velocity profile of the giant's wind as derived by 
\cite{d+99} (Fig.~\ref{fig:h1h2}), and calculated 
$n_{\rm H}(i,\phi)$ as a function of the orbital inclination 
and the phase (Figs.~\ref{fig:nhfi} and \ref{fig:nhi}). 
Third, comparing the $n_{\rm H}(i,\phi)$ function with the 
observed value of $3.9\times 10^{22}$\cmd, we estimated the 
range of orbital inclinations as follows: 
(i) for the average value of the parameter $X^{H+} = 17$, 
  $i = 59^{\circ} - 2^{\circ}\,/ + 3^{\circ}$, where the 
uncertainties reflect those in the binary position, and 
(ii) for the most probable position of the binary at 
$\phi = 0.961$, the range of $3 \la X^{H+} \la 30$ restricts 
the orbital inclination to 
  $i = 59^{\circ} - 2^{\circ}\,/ + 1^{\circ}$. 
Systematic errors given by using different wind velocity laws 
can increase $i$ up to $\sim 74^{\circ}$ (Sect.~4.1). 
Fluxes, that are measured around the inferior conjunction of 
the giant are consistent with the Rayleigh scattering light 
curves (Eq.~\eqref{eq:fray}) calculated for our resulting 
$n_{\rm H}(i,\phi)$ models (Fig.~\ref{fig:uvlc}).

The high value of the orbital inclination is also supported 
by the large amplitude of the orbitally related variation 
in the far-UV continuum (Sect.~4.2, Fig.~\ref{fig:uvlc}) 
and by the obscuration of the \ion{O}{i}]\,1641\,\AA\ fluxes 
around the inferior conjunctions of the giant during 
the quiescent phase of Z~And (Sect.~4.3, Fig.~\ref{fig:oi]}). 

Finally, the higher value of the inclination of the Z~And orbital 
plane allows us to interpret the satellite components of \ha\ and 
\hb\ emission lines, which appeared during active phases from 2006 
as highly collimated jets rather than as a special type of 
radiatively accelerated wind. 

\begin{acknowledgements}
We thank the anonymous referee for constructive comments. This 
research was supported by a grant of the Slovak Academy of 
Sciences, VEGA No. 2/0038/10, and by the realization of the 
Project ITMS No. 26220120029, based on the supporting operational 
Research and development program financed from the European 
Regional Development Fund.
\end{acknowledgements}

\begin{thebibliography}{}
\bibliographystyle{aa}
%
\bibitem[Crowley et al. (2005)]{crowley+05}
         Crowley, C., Espey, B. R., \& McCandliss, S. R. 
         2005, In Proc. of the 13th Cambridge Workshop on Cool 
         Stars, Stellar Systems and the Sun, eds. F. Favata, 
         G. Hussain \& B. Battrick, (ESA SP-560: ESA), p. 343
%
\bibitem[Crowley et al. (2008)]{crowley}
         Crowley, C., Espey, B. R., McCandliss, S. R. 
         2008, ApJ, 675, 711
%
\bibitem[Dumm et al. (1999)]{d+99}
         Dumm T., Schmutz W., Schild H., Nussbaumer H., 
         1999, A\&A, 349, 169
%
\bibitem[Fern\'andez-Castro et al. (1988)]{f-c88}
         Fern\'andez-Castro, T., Cassatella, A., Gim\'enez, A. 
         Viotti, R. 1988, ApJ, 324, 1016
%
\bibitem[Fekel et al. (2000a)]{fekel+00a}
         Fekel, F. C., Joyce, R. R., Hinkle, K. H., \& Skrutskie, M.
         2000a, AJ, 119, 1375
%
\bibitem[Fekel et al. (2000b)]{fekel+00b}
         Fekel, F. C., Hinkle, K. H., Joyce, R. R., \& Skrutskie, M. 
         2000b, AJ, 120, 3255
%
\bibitem[Horne et al. (1994)]{ho+94}
         Horne, K., Marsh, T. R., Cheng, F. H., Huben\'y, I., 
         \& Lanz, T. 1994, ApJ, 426, 294
%
\bibitem[Isliker et al. (1989)]{inv89} 
         Isliker, H., Nussbaumer, H., \& Vogel, M.
         1989, A\&A, 219, 271
%
\bibitem[Isogai et al. (2010)]{isogai+10}
         Isogai, M., Seki, M., Ykeda, Y., Akitaya, H., 
         Kawabata, K.S., 2010, AJ, 140, 235
%
\bibitem[Kilpio et al. (2011)]{kilpio+11}
         Kilpio, E., Bisikalo, D., Tomov, N., \& Tomova, M. 
         2011, Ap\&SS, 335, 155
%
\bibitem[Miko\l ajewska et al. (1995)]{mika+95}
         Miko\l ajewska, J., Kenyon, S. J., Miko\l ajewski, M., 
         Garcia, M. R., \& Polidan, R. S. 
         1995, AJ, 109, 1289
%
\bibitem[Miko\l ajewska \& Kenyon (1996)]{mk96}
         Miko\l ajewska, J., Kenyon, S. J., 1996, AJ 112, 1659
%
\bibitem[M\"urset et al. (1991)]{m+91}
         M\"urset, U., Nussbaumer, H., Schmid, H. M., \& Vogel, M.
         1991, A\&A, 248, 458 
%
\bibitem[M\"urset \& Schmid (1999)]{m+s99}
         M\"urset, U., \& Schmid, H. M. 1999, A\&AS, 137, 473
%
\bibitem[Nussbaumer \& Vogel (1987)]{nv87}
         Nussbaumer, H., Vogel, M. 1987, A\&A, 182, 51
%
\bibitem[Nussbaumer \& Vogel (1989)]{nv89}
         Nussbaumer, H., \& Vogel, M. 1989, A\&A, 213, 137
%
\bibitem[Nussbaumer et al. (1989)]{nsv89} 
         Nussbaumer, H., Schmid, H. M., \& Vogel, M. 
         1989, A\&A, 211, L27
%
\bibitem[Pereira et al. (1999)]{pe99}
         Pereira, C. B., Ortega, V. G., Monte-Lima, I. 
         1999, A\&A, 344, 607
%
\bibitem[Reimers (1981)]{reimers}
         Reimers, D. 1981, in Physical Processes in Red Giants, 
         ed. I. Iben and A. Renzini (Dordrecht: Reidel), p. 269
%
\bibitem[Schmid \& Schild (1997a)]{ss97a}
         Schmid, H. M., Schild, H. 1997a, A\&A, 327, 219
%
\bibitem[Schmid \& Schild (1997b)]{ss97b}
         Schmid, H. M. \& Schild, H. 1997b, A\&A, 321, 791
%
\bibitem[Seaquist et al. (1984)]{stb}
         Seaquist, E. R., Taylor, A. R., Button, S. 
         1984, ApJ, 284, 202 (STB)
%
\bibitem[Shore \& Aufdenberg (1993)]{sa93}
         Shore, S. N., \& Aufdenberg, J. P. 
         1993, ApJ, 416, 355
%
\bibitem[Shore \& Wahlgren (2010)]{s+w10}
         Shore, S. N., \& Wahlgren, G. M. 
         2010, A\&A, 515, A108
%
\bibitem[Skopal (2003)]{sk03}
         Skopal, A. 2003, A\&A, 401, L17
%
\bibitem[Skopal (2005)]{sk05}
         Skopal A., 2005, A\&A 440, 995
%
\bibitem[Skopal \& Pribulla (2006)]{sp06}
         Skopal, A. \& Pribulla, T. 
         2006, ATel. No. 882 
%
\bibitem[Skopal et al. (2009)]{sk+09}
         Skopal, A., Pribulla, T., Budaj, J., et al. 
         2009, ApJ, 690, 1222
%
\bibitem[Sokoloski et al. (2006)]{sok+06}
         Sokoloski, J. L., Kenyon, S. J., Espey, B. R., et al. 
         2006, ApJ, 636, 1002
%
\bibitem[Tomov et al. (2010)]{tomov+10}
         Tomov, N., A., Bisikalo, D. V., Tomova, M. T., 
         \& Kil'pio, E. Yu. 2010, Astron. Rep., 54, 628
%
\bibitem[Vogel (1991)]{vo91}
         Vogel M., 1991, A\&A, 249, 173
%
\bibitem[Vogel et al. (1992)]{v+92}
         Vogel, M., Nussbaumer, H., \& Monier, R. 
         1992, A\&A, 260, 156
%
\end{thebibliography}
\end{document}